\documentclass[aps,prl,twocolumn,amsfonts]{revtex4} 
\usepackage{epsfig,amsopn}
\begin{document}

\title{A Mechanism for Polar Protein Localization in Bacteria}

\author{Martin Howard}

\affiliation{
Department of Mathematics, Imperial College London, South
Kensington Campus, London SW7 2AZ, U.K.}

\begin{abstract}
We investigate a mechanism for the polar localization of proteins in
bacteria.  We focus on the MinCD/DivIVA system regulating division
site placement in the rod--shaped bacterium {\em Bacillus
subtilis}. Our model relies on a combination of geometric effects and
reaction--diffusion dynamics to direct proteins to both cell poles,
where division is then blocked. We discuss similarities and
differences with related division models in {\em Escherichia coli} and
also develop extensions of the model to asymmetric polar protein
localization. We propose that our mechanism for polar localization may
be employed more widely in bacteria, especially in outgrowing spores,
which do not possess any pre--existing polar division apparatus from
prior division events.
\end{abstract}

\date{\today} \maketitle

\section{Introduction}

The polar localization of proteins is a vital component of the
sophisticated internal organization of bacteria, and is important for
many processes ranging from chemotaxis to accurate cell division
\cite{Shapiro,Errington,Lutkenhaus,Shapiro2}. However, how such
polar localization is achieved has remained largely unknown. The
purpose of this report is to analyze one possible mechanism for
localization in a well--studied experimental context: the MinCD/DivIVA
system regulating cell division positioning in the Gram--positive
bacterium {\em B. subtilis} \cite{Cha,Edwards1,
Marston1,Marston2,Karoui,Muchova,Harry,Hamoen}. We will then discuss
how these ideas might be applied more generally to the generation of
polar localization/asymmetry in bacteria.

The dynamics of bacterial cell division have been intensively studied,
particularly in {\em B. subtilis} \cite{Cha,Edwards1,
Marston1,Marston2,Karoui,Muchova,Harry,Hamoen} and {\em E. coli} 
\cite{Raskin1,Raskin2,Hu1,Yu,Rowland,Fu,Rothfield,Hale,Hu2,Suefuji,
Shih1},
where the polar localization of proteins is known to be
crucial. In both cases, the location of the division site is
determined by the tubulin--like GTPase FtsZ which assembles into a
ring--shaped structure just under the cell membrane at midcell. In
{\em E. coli} accurate targeting of the FtsZ ring to midcell is
achieved both by the inhibitory effect of the nucleoids (``nucleoid
occlusion'') and through the MinCDE proteins \cite{Yu}. MinC and MinD
form a complex on the cell membrane, which undergoes coherent
pole--to--pole oscillations, with a period of around a minute, driven
by MinE. As a result of the oscillations, the time--averaged MinCD
concentration is minimized at midcell and maximized at the cell
poles. Since MinC inhibits FtsZ polymerization, the FtsZ ring
assembles at midcell where the average MinC concentration is
lowest. In this way, through the temporary sequestration of MinCD at
the cell poles, the midcell can be accurately targeted for cell
division.

Recently, several groups have employed simple reaction--diffusion
mathematical models (either deterministic
\cite{Howard1,Meinhardt,Kruse} or stochastic \cite{Howard2}) to try to
understand the MinCDE dynamics (see also
Refs.~\cite{Meinhardt2,Hunding} for 
related models). The principal result of this modeling effort
is that the oscillations are a spontaneous self--organized phenomena
arising from an intrinsic instability of the reaction--diffusion
dynamics. A vital element of this instability is the varying diffusion
constants of the MinCDE proteins: high in the cytoplasm; low when
polymerically bound to the membrane. Crucially, the oscillations form
spontaneously as a result of this disparity of the diffusion constants
and of the nonlinear binding/unbinding interactions between the
proteins. As a result, the oscillatory dynamics does not require any
pre--existing topological markers to distinguish various locations
within the cell, as had previously been supposed.

Very recently, improved fluorescent imaging has revealed that the {\em
E. coli} Min proteins polymerize into stable helices on the
cytoplasmic membrane \cite{Shih1}. This phenomenon is not explicitly
included in any of the existing mathematical models, which may
therefore require some amendment. Nevertheless, the models generally
agree well with fluorescent imaging experiments, especially in the
observation of periodic MinCDE/FtsZ bands in filamentous
mutants. These results are very difficult to understand other than as
a pattern--forming reaction--diffusion instability. Hence, we believe
the essentials of a reaction--diffusion mechanism underpinning the
MinCDE oscillations are likely to be correct. Although the formation
of the Min proteins into filaments has only been observed in {\em
E. coli} (and also {\em in vitro} \cite{Suefuji}), this observation
may have important implications in other bacteria. Already in {\em
B. subtilis} the shape--determining proteins MreB and Mbl are known to
form helical cables that could comprise cytoskeletonal--like
elements \cite{Errington,Jones}.  

Despite possessing two of the three same proteins (MinC and MinD, but
not MinE), {\em B. subtilis} appears to use a quite different
mechanism to achieve central placement of the FtsZ ring in vegetative
cells
\cite{Cha,Edwards1,Marston1,Marston2,Karoui,Muchova,Harry,Hamoen}. In
{\em B. subtilis}, MinCD do not oscillate, but instead are localized
to both cell poles, where they are anchored by the protein DivIVA
(which is structurally unrelated to MinE).  Also involved are the
(initially) homogeneously distributed proteins ZapA, which
promotes \cite{Gueiros}, and EzrA, which inhibits \cite{Levin1}, FtsZ ring
assembly.  Quite how DivIVA is able to recognize the cell poles has
remained unresolved and forms the main object of this report. With
MinCD localized to the cell poles, the FtsZ ring is again unable to
nucleate there. However, it is likely that additional mechanisms are
used to accurately locate the FtsZ ring to midcell \cite{Levin2},
since the MinCD/DivIVA system functions primarily to block polar
division \cite{Migocki}. These additional (and so far uncharacterized)
mechanisms may be especially important as nucleoid occlusion is
ineffective as a means of positioning in {\em B. subtilis} \cite{Regamey}.
Despite this, the precision of midcell division is the
same in {\em E. coli} and {\em B. subtilis} \cite{Migocki}, even
though their regulatory mechanisms differ considerably.

Polar protein localization is not only restricted to positioning of
the division site. For example in {\em Caulobacter crescentus}, polar
asymmetry is exploited to ensure that daughter cells are dissimilar:
one a swarmer cell with a polar flagellum, and one a stalked cell with
a polar stalk that can adhere to a source of food
\cite{Shapiro,Shapiro2}. In
each case, the development of a cell with a polar stalk or flagellum
is critically dependent on spatiotemporally controlled protein
localization, and on the subsequent proteolysis and release of these
components at the appropriate time. Other examples of polar
localization include the polar clustering of chemoreceptors, in both
{\em E. coli} \cite{Maddock,Lybarger,Maki} and {\em C. crescentus}
\cite{Alley}, and also the polar
localization of the virulence protein
IcsA \cite{Robbins}, which promotes undirectional actin--based
motility of the bacterium {\em Shigella flexneri}.

\section{Polar protein localization in {\em B. subtilis}}

Polar division in vegetative cells of {\em B. subtilis} is blocked by
the presence of membrane localized MinC. MinC is normally a
cytoplasmically localized protein, however, in the presence of MinD
the two interact strongly and bind to the membrane
\cite{Marston2}. Polar localization is facilitated by the action of
the protein DivIVA, which acts to anchor MinCD to the poles
\cite{Marston1,Marston2}. Targeting of MinCD to the poles, is
dependent on DivIVA, as in filamentous cells lacking DivIVA, MinCD was
partly distributed diffusively along the cell
\cite{Marston1,Marston2}. Furthermore the localization and activity of
MinD is believed to require ATP binding and possibly hydrolysis
\cite{Karoui}.  However, MinCD can also (MinC more strongly than MinD)
target past/present division locations independently of DivIVA
\cite{Marston1,Marston2}. Similarly in a {\em minD} mutant, DivIVA
simply localizes to division sites regardless of location, again
demonstrating an affinity for some component of the cell division
apparatus \cite{Marston1}.  Thus in wild--type cells, MinCD/DivIVA
localize to the poles (old division sites), blocking cell division
there. However, before septal constriction occurs and the cell
divides, MinCD/DivIVA assemble approximately simultaneously at midcell
\cite{Marston1}, and are retained there after division
\cite{Marston1,Marston2}. In this way MinCD is delivered to the new
poles of the two daughter cells, meaning that polar division is
immediately prevented at the new poles.

This observation immediately provides one possible explanation for
polar localization: since the cell division apparatus marks the
location of the new cell poles in the daughter cells, if the relevant
proteins can localize to a component of the division apparatus, then
the cell pole will continue to be marked in the daughter
cells. However, one may question whether this mechanism is really
sufficiently robust: if MinCD/DivIVA fail to correctly localize to the
site of division, then the poles of the resulting daughter cells will
remain unmarked, meaning that division in those cells would be
corrupted. It would certainly therefore be desirable to have other
means of identifying the cell poles.  Experiments in outgrowing spores
of {\em B. subtilis} indicate that this is indeed the case
\cite{Harry,Hamoen}. MinCD/DivIVA are targeted to the cell poles very
rapidly after spore germination, even in cells lacking FtsZ. Thus,
polar localization can occur even in germinating spores which do not
possess pre--existing division apparatus left over from prior division
events.

So the question arises, what other mechanism could be responsible for
the polar localization? One possible answer is that a self--organized
reaction--diffusion instability exists, similar to that already
studied in {\em E. coli}. Of course, for {\em B. subtilis}, the
instability would have to be to a steady--state distribution where the
MinCD/DivIVA do not oscillate (i.e. in linear stability analysis one
would expect a real rather than a complex growth eigenvalue). With
appropriate interactions and diffusion constants, such an instability
could certainly exist. However, one of the key predictions of such an
instability is the existence of a characteristic wavelength, depending
on the microscopic parameters of the model, but independent of the
geometry \cite{Howard1,Meinhardt,Kruse}. Thus in long filamentous
cells of {\em E. coli}, several wavelengths can be fit into a single
cell and one therefore expects to see multiple oscillating bands. The
observation of this phenomena in filamentous {\em E. coli}
\cite{Raskin1} is powerful evidence in favor of the
reaction--diffusion mechanism. The situation in {\em B. subtilis} is,
however, rather different: in filamentous cells lacking FtsZ,
MinCD/DivIVA are primarily localized to the cell poles
\cite{Marston1,Marston2}, without any regularly spaced intermediate
bands in between, as one would expect for a reaction--diffusion
instability. One could argue that the affinity of MinCD/DivIVA for the
pre--existing (polar) division apparatus from previous division events
might mask the existence of any intermediate bands.
However, in outgrowing spores, which have no pre--existing division
apparatus, there are still no intermediate bands, even in long
filamentous mutants (up to about $13~\mu m$ long) \cite{Harry,Hamoen}.
Thus we see no evidence for the existence of a
characteristic wavelength. Of course, it is just possible that the
characteristic wavelength is simply very long, too long to have been
seen in past experiments; but, there is currently no hint of this.

Instead we propose that it is a geometric effect in combination with
reaction--diffusion dynamics which marks the end of the cell. The
essence of our approach is to assume that MinCD has a {\em reduced}
rate of membrane binding at the cell poles. If, for example, MinCD
polymerize into filaments, as in {\em E. coli} \cite{Shih1}, this
might be due to the increased curvature of the cell poles interfering
with membrane polymerization. Since DivIVA may then bind to the edges
of MinCD clusters and so stabilize them, the MinCD (and DivIVA)
concentrations end up being maximized close to the cell poles, even
though MinCD initially has a reduced rate of binding there.
Similarly, when the cell divides a region of high curvature is created
at midcell, again attracting the MinCD/DivIVA proteins independent of
the division apparatus. In the next section, we define
this model more precisely and report on the results of numerical
simulations.

\section{The model}

Previous modeling work on the MinCDE system has demonstrated the
effectiveness of simple deterministic reaction--diffusion equations
\cite{Howard1,Meinhardt,Kruse}. Accordingly we follow a similar
modeling approach here. The dynamics of MinC is similar to that of
MinD, so, for simplicity, we drop MinC from our model. The fact that
MinD can localize normally in {\em minC} mutant cells also supports
this approach \cite{Marston2}. Initially, we also assume a
``conservative'' model, where the MinD/DivIVA are able to bind/unbind
from the membrane but are not degraded in the process. This is again
similar to some of the models used for {\em E. coli}
\cite{Howard1,Kruse,Howard2}. The relevant set of one--dimensional
reaction--diffusion equations are:
\begin{eqnarray}
& & \hspace{-0.4cm} {\partial\rho_D\over\partial t}= D_D
  {\partial^2\rho_D\over\partial x^2}-{\lambda_D(x)\rho_D\over
  (1+\lambda_D'^2\rho_d^2)}+{\lambda_d\rho_d\over
  (1+\lambda'_d\rho_a)} \nonumber \\
& & \hspace{-0.4cm} {\partial\rho_d\over\partial t}= \hspace{0.15cm}D_d
  {\partial^2\rho_d\over\partial x^2}\hspace{0.1cm}+{\lambda_D(x)\rho_D\over
  (1+\lambda_D'^2\rho_d^2)}-{\lambda_d\rho_d\over
  (1+\lambda'_d\rho_a)} \nonumber \\
& & \hspace{-0.4cm} {\partial\rho_A\over\partial t}= \hspace{0.05cm} D_A
  {\partial^2\rho_A\over\partial x^2}
  -{\lambda_A\rho_A\left|{\partial\rho_d\over\partial
  x}\right|\over (1+\lambda_A'^2\rho_a^2)}+ {\lambda_a\rho_a\over
  (1+\lambda_a'^2\rho_a^2)} \nonumber \\
& & \hspace{-0.4cm} {\partial\rho_a\over\partial t}= \hspace{0.15cm} D_a
  {\partial^2\rho_a\over\partial x^2}\hspace{0.1cm}
  +{\lambda_A\rho_A\left|{\partial\rho_d\over\partial
  x}\right|\over (1+\lambda_A'^2\rho_a^2)}- {\lambda_a\rho_a\over
  (1+\lambda_a'^2\rho_a^2)} \nonumber ~ .
\end{eqnarray}
Here, the first equation is for the cytoplasmic MinD density $\rho_D$,
the second for the membrane bound MinD density $\rho_d$, the third for
the cytoplasmic DivIVA density $\rho_A$, and the fourth for the
membrane bound DivIVA density $\rho_a$.  In the above model, we assume
that MinD binds to the membrane with rate $\lambda_D$, in agreement
with green fluorescent protein (GFP) fusions which indicate that MinD
is a membrane--localized protein \cite{Marston1, Marston2}. However,
one would also expect that the rate of binding is reduced as the MinD
membrane density increases. This accounts for the $\lambda_D'$
term. We also assume that MinD can unbind spontaneously ($\lambda_d$
term), but that this is suppressed by DivIVA ($\lambda_d'$ term)
\cite{Marston1,Marston2}. It is also possible to insert a factor
$(1+\lambda_d''\rho_d)^{-1}$ in this term, which would model increased
stabilization of membrane MinD from clustering. However, we have found
that our results were not significantly affected by this term, which
we have therefore left out for simplicity. Notice that the DivIVA
functions to stabilize the membrane MinCD, hence its role is very
different from the {\em E. coli} MinE protein which destabilizes
membrane MinCD. This difference is partly responsible for the
structure of the equations being rather different in this case from
the earlier models for {\em E. coli} MinCDE
\cite{Howard1,Meinhardt,Kruse}. DivIVA is also believed to be an
intrinsically cytoplasmic protein \cite{Cha}. Hence, in order to bind
to the membrane in outgrowing spores \cite{Harry}, (without
pre--existing division apparatus), it presumably requires the presence
of MinD. We assume here, similar to Ref.~\cite{Karoui}, that it binds
to (and then stabilizes) the edges of MinD clusters. This could
correspond to DivIVA ``capping'' the ends of MinCD filaments
\cite{Karoui}. This process would then account for
the $\lambda_A$ gradient term $|\partial\rho_d/\partial x|$, which is
maximized at the edges of a cluster. Note that the gradient term has a
modulus sign, as the rate of binding should not depend on the sign of
the gradient. 
Again we assume that the DivIVA binding is suppressed as
the membrane DivIVA density increases ($\lambda_A'$ term). Finally
DivIVA can spontaneously unbind ($\lambda_a$ term), but this is
suppressed by DivIVA itself ($\lambda_a'$ term).  This indicates that
DivIVA itself oligomerizes/forms clusters for maximum stability, for
which there is also experimental evidence \cite{Muchova}.

The above assumptions are fairly minimal and are all closely based on
the available experimental evidence of GFP fusions and deletion
mutants \cite{Cha,Edwards1,
Marston1,Marston2,Karoui,Muchova,Harry,Hamoen}. As we have said, we
then add one extra ingredient, namely a position dependent membrane
binding $\lambda_D(x)$ for the MinD, where $\lambda_D(x)$ is reduced
in regions close to the cell ends (see below).

For our computer simulations, we assume hard wall boundary conditions
(i.e. $\partial\rho_i/\partial x=0,~~i=\{D,d,A,a\},$ at the
boundaries) and we numerically integrate the above equations, with
lattice size $dx=0.02~\mu m$ and time increment $dt=1\times
10^{-5}~s$. We start from random initial conditions, where the MinD
and DivIVA are randomly distributed throughout the cytoplasm. We begin
the simulation with a total of $50$ lattice sites (bacterial length
$1~\mu m$), with initial densities of $1000 ~\mu m^{-1}$ for MinD and
$750~\mu m^{-1}$ for DivIVA. The actual concentration levels for {\em
B. subtilis} have not been accurately measured, so we assume
concentration levels similar to those observed in {\em E. coli}
\cite{Shih2}.  During the course of the simulation, we then allow the
bacterial length to increase, modeling a growing bacterium, with
growth rate of $4~\mu m/{\rm hour}$. In other words, every $1.8\times
10^5$ time steps an extra site is inserted, randomly located. Note,
however, that the extra site is never inserted at the inert cell ends
\cite{Daniel} (defined here by regions where $\lambda_D(x)$ is
reduced, see below). As the bacteria grows in the simulation we add
extra MinD/DivIVA, randomly localized in the cytoplasm, to ensure that
the average densities remain constant (i.e. the total number of
MinD/DivIVA proteins is proportional to the cell length). There is
currently no information on whether the average densities do actually
remain constant as the cell grows, but the assumption used here is the
simplest possible. Nevertheless, precise regulation of the densities
turns out not to be particularly important. Considerable
overproduction eventually leads to the MinCD/DivIVA proteins
saturating the poles and spilling over into the reminder of the cell
length. Underproduction simply translates to less MinCD/DivIVA being
present along the cell length in between the poles. The important
point is that the bacterium has to manufacture sufficient MinCD/DivIVA
to ensure that the cell poles are always well populated with these
proteins, but not so much that the proteins spill over to midcell
leading to division inhibition. Note that in our simulations, we
exclusively model the case of outgrowing spores without pre--existing
division apparatus. Clearly, modeling protein localization to
pre--existing binding sites in wild--type vegetative {\em B. subtilis}
cells is straightforward, and not particularly revealing.

In our simulations we use the following parameter set: $D_D=2.0~\mu
m^2 s^{-1}$, $D_d=0$, $D_A=2.0~\mu m^2 s^{-1}$, $D_a=2\times
10^{-4}~\mu m^2 s^{-1}$, $\lambda_D=10~s^{-1}$, $\lambda_D'=0.001~\mu
m$, $\lambda_d=3.0~s^{-1}$, $\lambda_d'=0.2~\mu m$,
$\lambda_A=0.03~\mu m^2 s^{-1}$, $\lambda_A'=8\times 10^{-4}~\mu m$,
$\lambda_a=2.0~s^{-1}$, $\lambda_a'=0.005~\mu m$. Furthermore, for the
closest $10$ sites to both ends of the bacteria, the binding rate
$\lambda_D$ is reduced by a factor of $2$ (our results are robust to
changes in this suppression factor).  It is also possible to
generalize the model so that the MinD/DivIVA proteins are degraded at
a (slow) rate, whilst increasing the production rate to keep the
average densities constant. We have verified that this modification
also makes no qualitative difference to our results. The values for
the above cytoplasmic diffusion constants are of the same order as
that measured (for different proteins) in the cytoplasm of {\em
E. coli} \cite{Elowitz}. As for the membrane diffusion constants, we
have used similarly small values to those used in {\em E. coli}
\cite{Howard1,Howard2}. Previously this was an important component in
the spontaneous reaction--diffusion instability; this is, however, not
the case here: the small membrane diffusivities are now necessary to
preserve the polar density concentration maxima and to prevent the
concentrations from equalizing everywhere along the cell membrane.
For the remaining parameters there is no experimental guidance, but we
have taken care to check that our results are robust to variations in
these parameters.

\section{Results: polar protein localization}

In Figure~1, we show the result of our computer simulation of the
spore outgrowth experiment. We see very similar results to those
observed experimentally \cite{Harry,Hamoen}. Starting from random
initial conditions, the MinD/DivIVA rapidly (within a few minutes)
localize to the poles of the cell. This agrees with experiments
indicating that MinD/DivIVA localize to the cell poles very early in
spore germination \cite{Harry,Hamoen}. In our model the mechanism for
this localization follows from the reduced rate of MinD binding to the
cell ends. This reduced polar value for $\lambda_D(x)$ is absolutely
critical for polar localization as it sets up a pronounced membrane
MinD density gradient close to the cell poles. The DivIVA then binds
to these positions of increased gradient, stabilizing the
MinD/DivIVA. In this way the MinD/DivIVA concentrations end up being
maximized close to the cell ends. In fact, the concentrations are not
maximized precisely at the poles, where there is actually a dip (see
Figures~2 and 3). However the polar concentrations still end up being
much higher than the densities well away from the cell ends. As the
cell continues to grow, the polar regions become saturated with
MinD/DivIVA and the newly produced proteins cannot localize
there. Instead the proteins localize to random membrane patches along
the cell length. Furthermore, these patches appear to propagate along
the cell length (see Figure~1). This movement results from the
coupling of the DivIVA to the {\em edges} of MinD clusters, slightly
away from the local MinD density maxima. Since MinD is stabilized by
the DivIVA this encourages MinD to rebind slightly away from its
current density maxima. This in turn shifts the position of the
maximum MinD density gradient, thus also encouraging the DivIVA to
rebind in a shifted location. In this way the coupled MinD/DivIVA
density maxima both tend to propagate along the membrane, as shown in
Figure~1. It would be interesting to see if such movement can be
observed in experiments. A snapshot of these patterns can be seen in
Figure~2, where we show the MinD/DivIVA membrane density profiles at
length $10~\mu m$. The DivIVA pattern is very similar to that observed
experimentally in filamentous spore outgrowth experiments
\cite{Harry,Hamoen}, with sharp polar peaks and less intense randomly
distributed membrane patches along the cell length. Note that this
disordered pattern is seen everywhere inside the simulated outgrowing
spore (including close to the cell ends) if $\lambda_D(x)$ is a
constant, emphasizing again that polar protein localization is
critically dependent on a position--dependent MinD binding rate.

\begin{figure}
\vbox{
\centerline{\hbox{
\hspace{0.0cm}
\epsfxsize=1.6in
\epsfbox{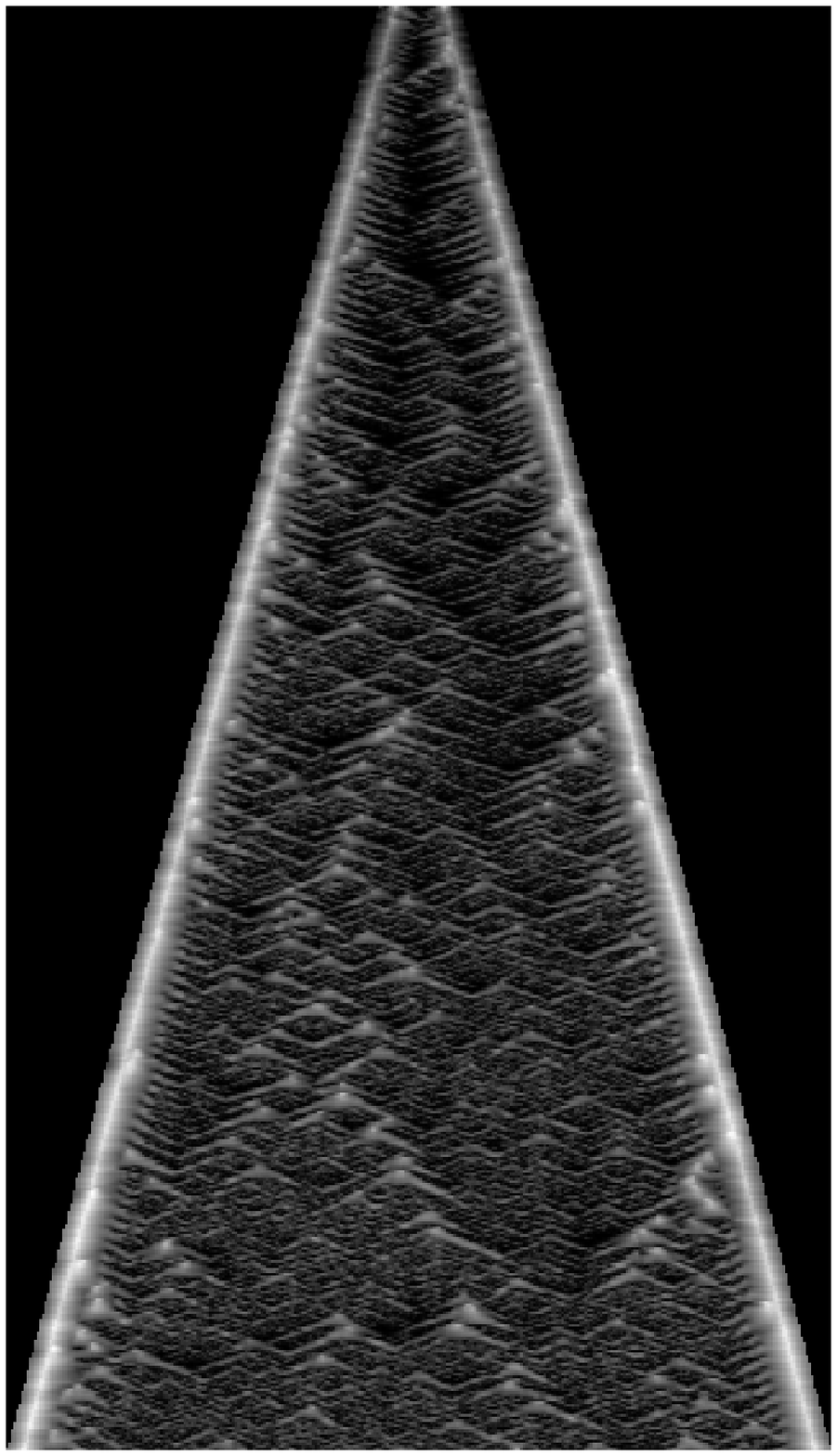}}
\hspace{0.1cm}
\epsfxsize=1.6in
\epsfbox{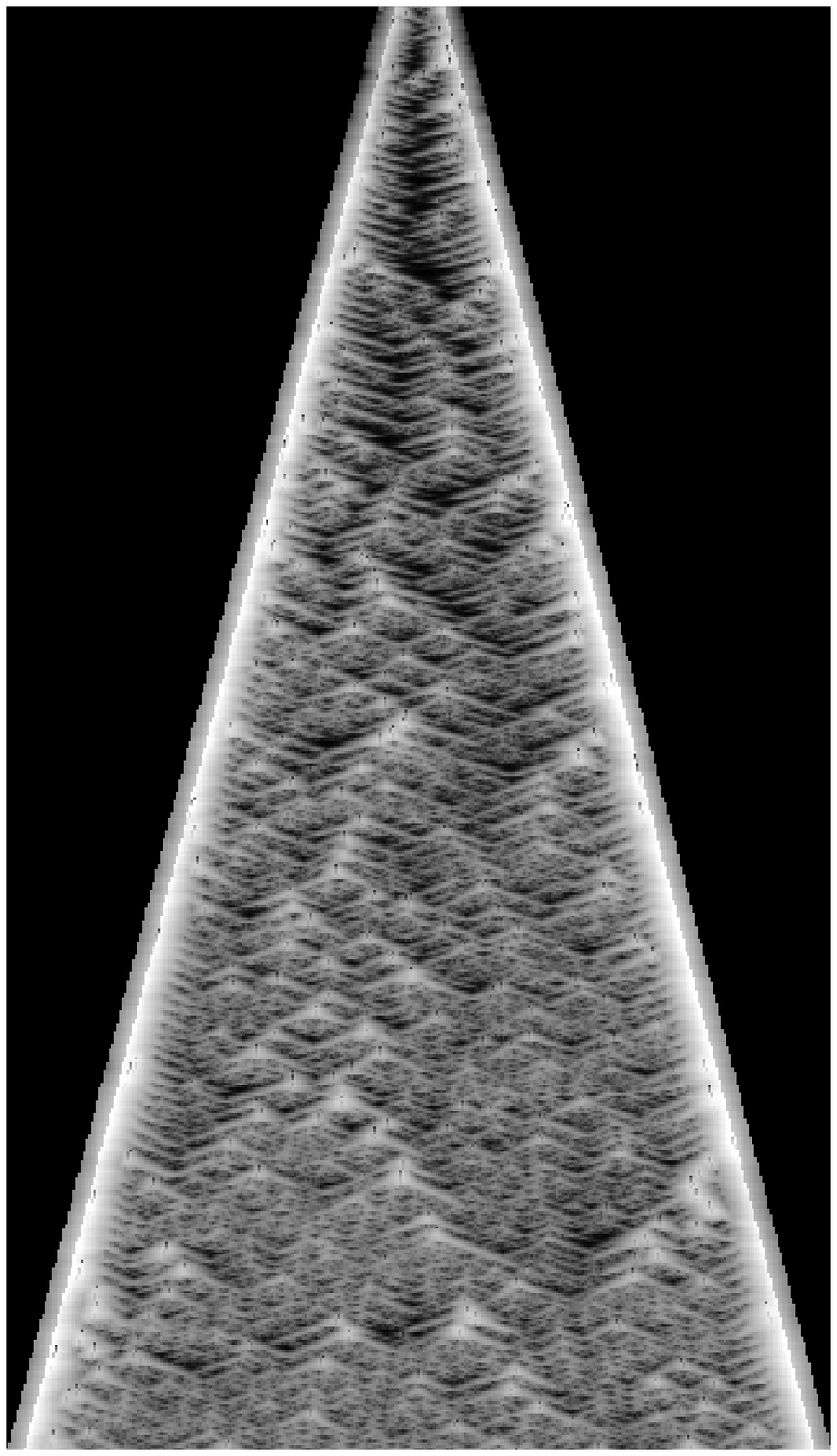}}
}
\caption{Spacetime plot of membrane--bound DivIVA (left) and MinD
  (right) densities in a simulated outgrowing spore. Time increases
  from top to bottom and the simulation lasts $8100~s$, starting with
  length $1~\mu m$ and ending with length $10~\mu m$. White (black)
  corresponds to areas of high (low) concentration.}
\label{fig1}
\end{figure}

\begin{figure}
\vbox{
\centerline{\hbox{
\epsfxsize=3.3in
\epsfbox{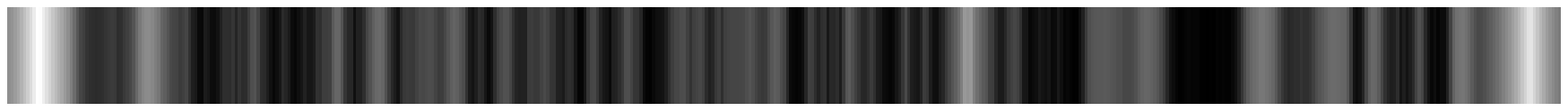}}
}\vspace{0.2cm}
\centerline{\hbox{
\epsfxsize=3.3in
\epsfbox{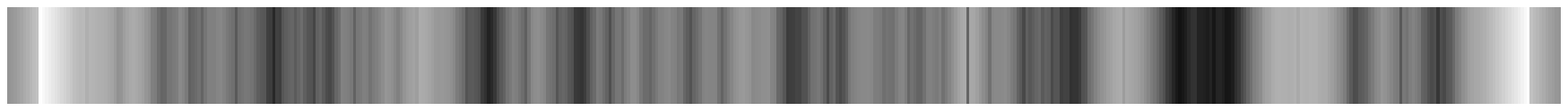}}
}
}
\caption{Snapshot of the DivIVA (top) and MinD (bottom) membrane
  density profiles at length $10~\mu m$. White (black) corresponds to
  areas of high (low) concentration. Note that these are one
  dimensional simulation results (two dimensional presentation in the
  figure is for clarity).}
\label{fig1+}
\end{figure}

More quantitative data, with profiles of the membrane MinD/DivIVA
densities at different times are shown in Figure~3(a,c,e). In
Figure~3(b,d,f) we show examples of the membrane density averaged over
$90~s$ as a function of position for various total cell lengths. A
$90~s$ average is chosen as, at least in {\em E. coli}, this is the
rough time over which the FtsZ averages in locating the FtsZ
ring \cite{Raskin1}. As can be seen, the MinD concentration is nicely
minimized at the cell center for shorter cell lengths. However, as the
cell length increases, the precision with which the center can be
identified is considerably reduced, with the averaged signal
fluctuating considerably at $8~\mu m$. This data is consistent with
evidence that the MinCD/DivIVA system is primarily responsible for
blocking polar division, rather than in accurately placing midcell
division \cite{Migocki}. Notice also how the MinD is less tightly
localized at the cell poles; its signal decreases more slowly than for
the DivIVA towards the center of the cell. This also agrees with
experiment \cite{Marston1}.

\begin{figure}
\vbox{
\centerline{\hbox{
\hspace{0.0cm}
\epsfysize=1.5in
\epsfbox{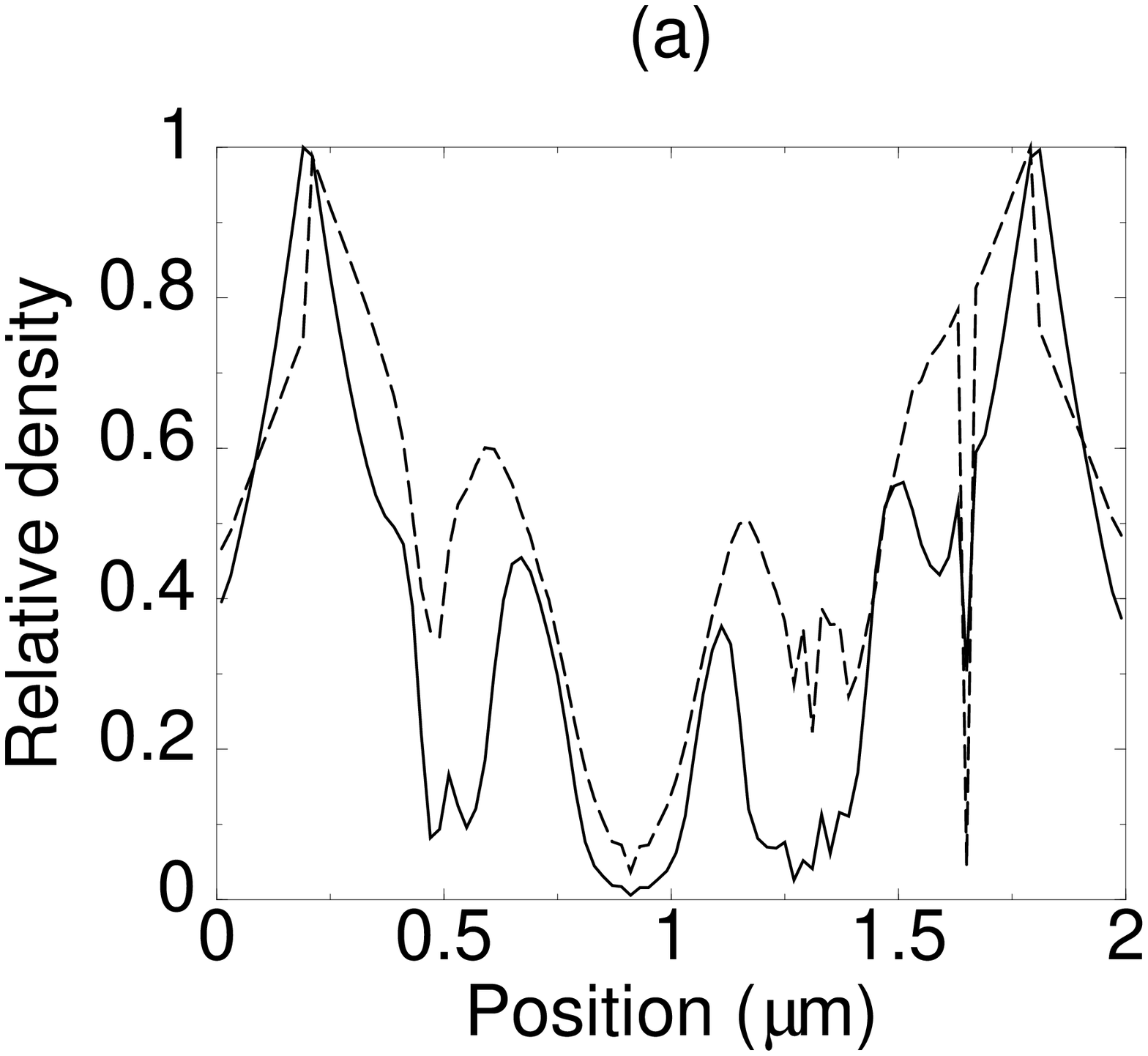}}
\hspace{0.2cm}
\epsfysize=1.5in
\epsfbox{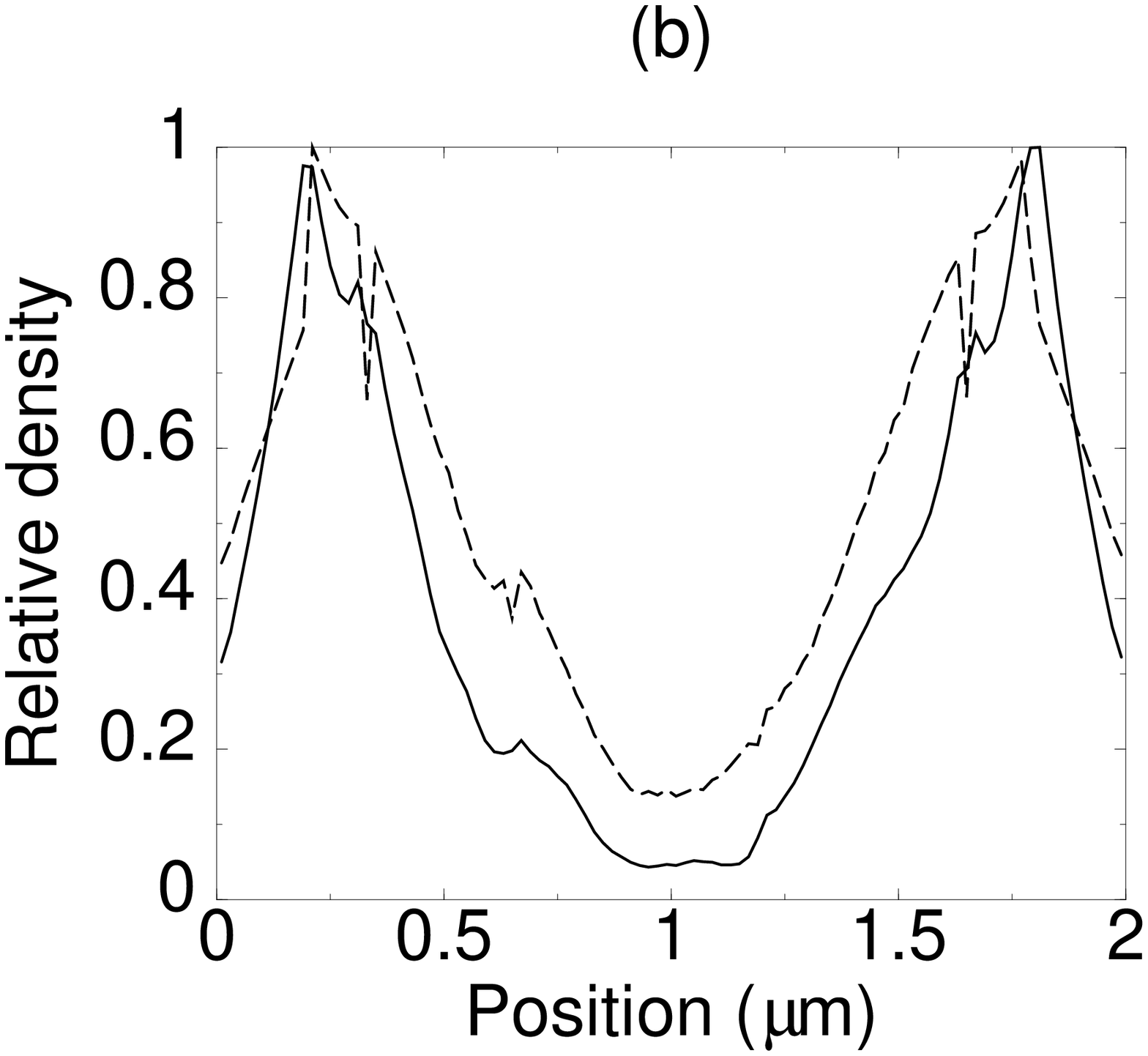}}
\vspace{.4cm}
\centerline{\hbox{
\hspace{0.0cm}
\epsfysize=1.5in
\epsfbox{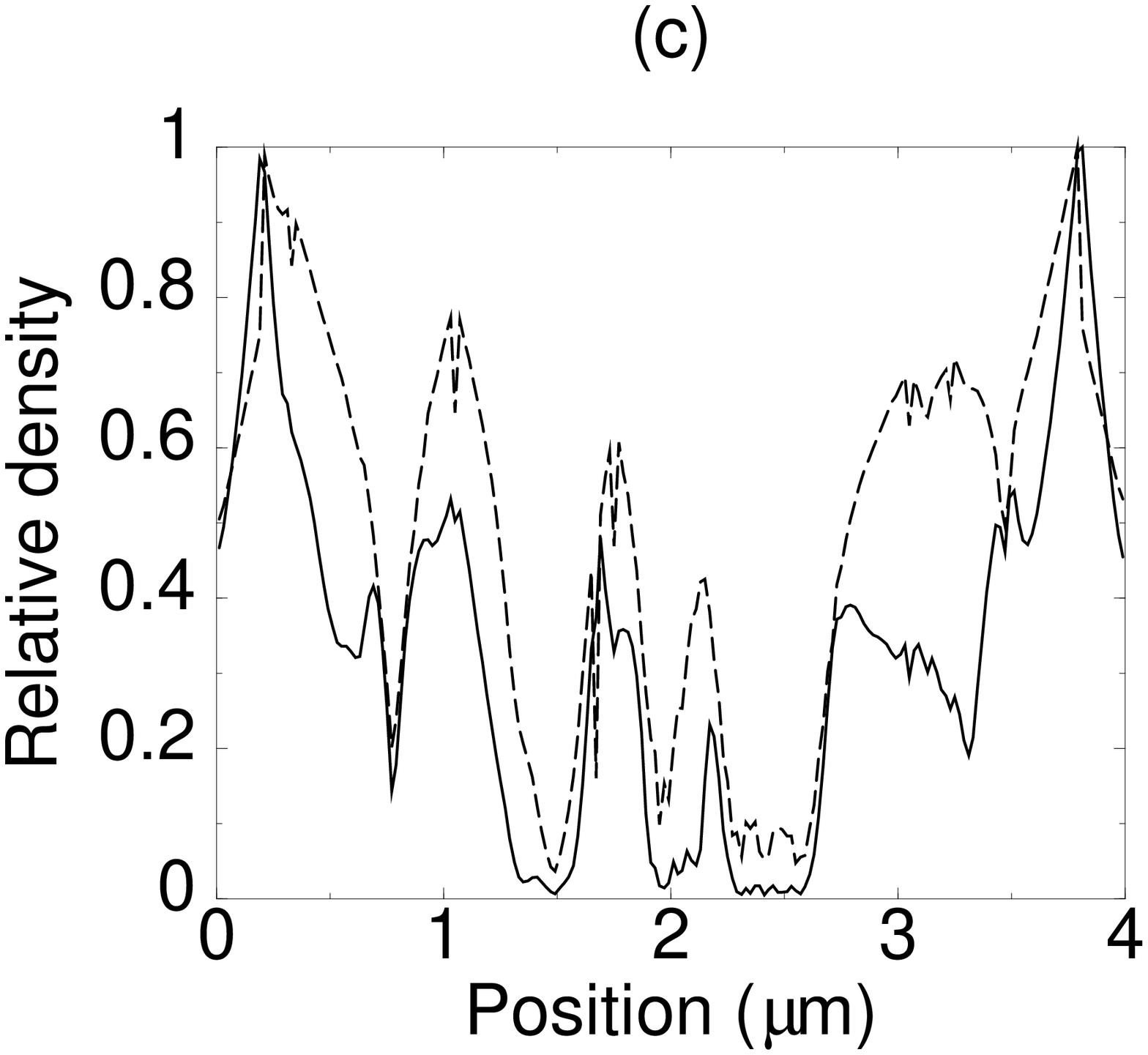}}
\hspace{0.2cm}
\epsfysize=1.5in
\epsfbox{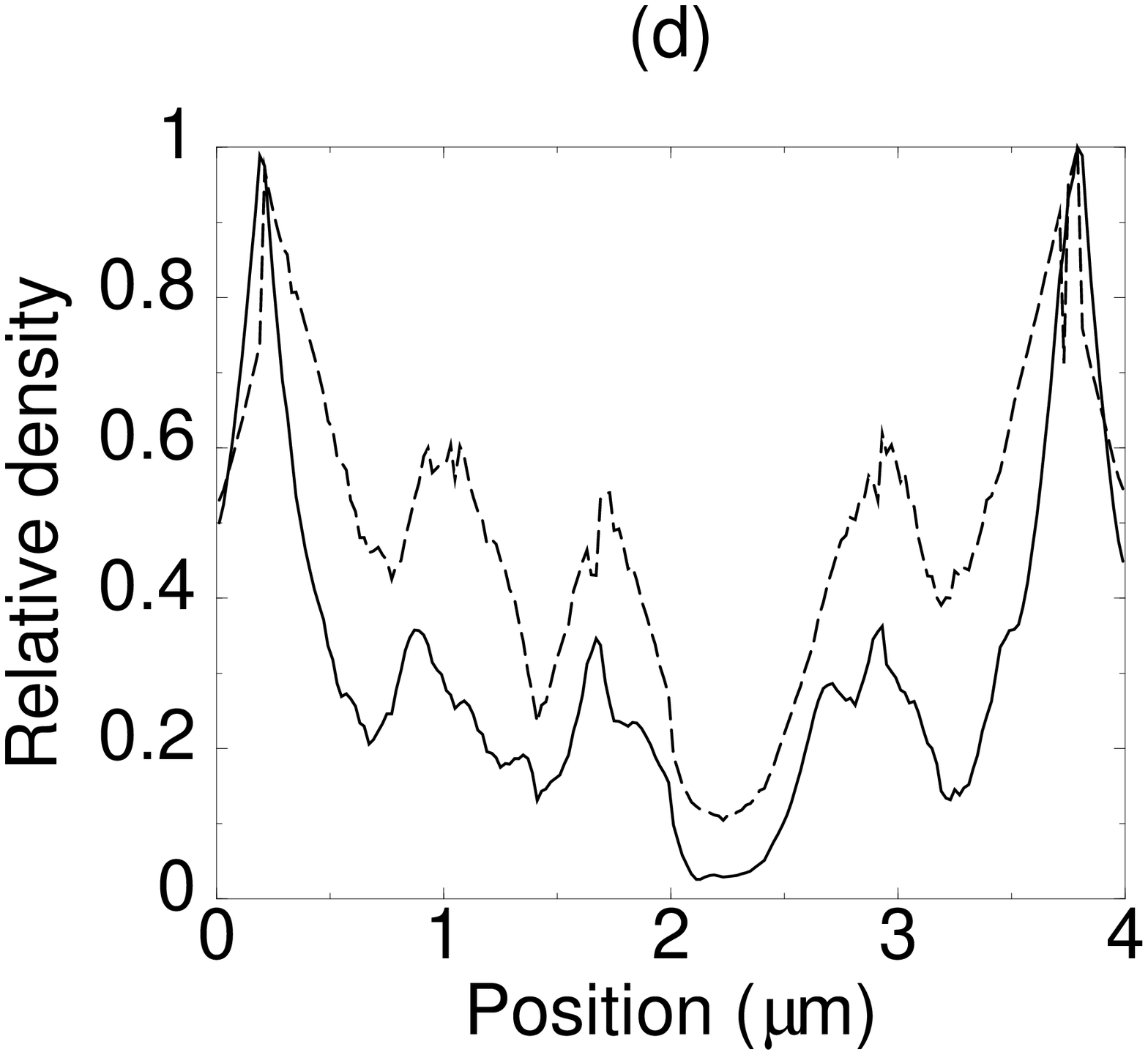}}
\vspace{.4cm}
\centerline{\hbox{
\hspace{0.0cm}
\epsfysize=1.5in
\epsfbox{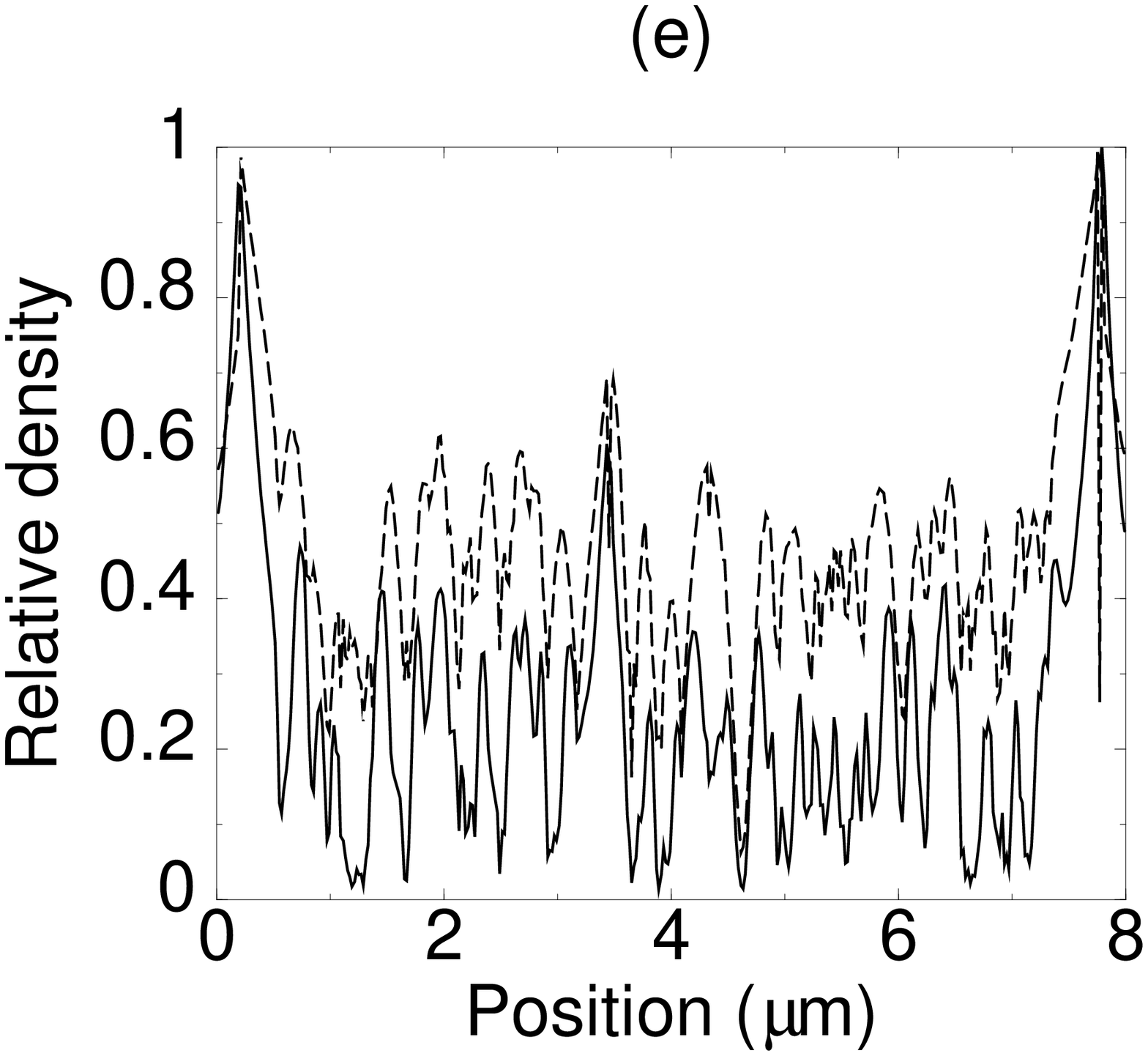}}
\hspace{0.2cm}
\epsfysize=1.5in
\epsfbox{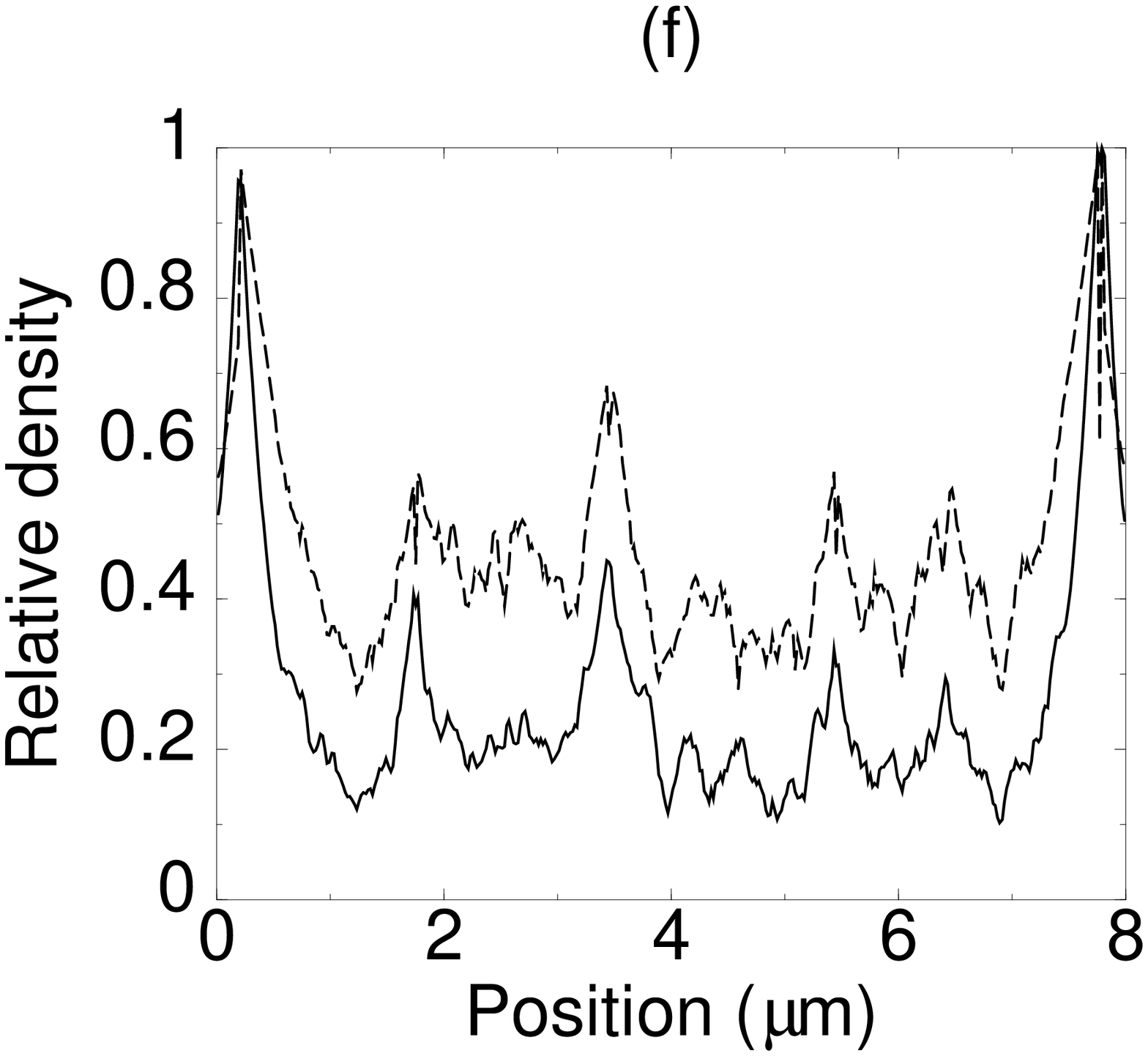}}
}
\caption{(a,c,e) Simulations of outgrowing spores: instantaneous
  membrane MinD (dashed line), and membrane DivIVA (full line) density
  profiles at lengths $2~\mu m$, $4~\mu m$ and $8\mu m$; (b,d,f)
  Membrane MinD (dashed line) and membrane DivIVA (full line) density
  profiles averaged over $90~s$ at lengths $2~\mu m$, $4~\mu m$ and
  $8~\mu m$. Maximum densities for each of the membrane MinD and
  DivIVA normalized to unity.  }
\label{fig2}
\end{figure}

\section{Results: asymmetric protein localization}

Up to this point, we have modeled symmetric protein localization, such
as that seen in the MinCD/DivIVA system in {\em B. subtilis}. However,
it is also interesting to ask whether one can modify the model to
obtain asymmetric localization, i.e. to one pole only. This
modification turns out to be quite straightforward: in essence one
merely has to remove the $\lambda_D'$ and $\lambda_A'$ terms which
restrict membrane binding when the membrane densities are already
high. In this case, starting from random initial conditions, the
dynamics rapidly localizes the proteins to both cell poles. However,
one randomly chosen pole will, due to fluctuations in the random
initial conditions, gain a slightly higher density than the
other. Since higher densities yield greater stability for the membrane
aggregations (due to the $\lambda_d'$ and $\lambda_a'$ terms), and
since there are no longer saturation effects, the favored pole, with
an initially slightly higher concentration, will continue to attract
additional proteins from the other pole, via cytoplasmic diffusion and
then polar membrane capture (cf. Ref.~\cite{Rudner}).  Of course this
process will also occur in reverse from the favored to the less
favored pole, but because of its higher density, the net rate of
accumulation will be towards the pole with the higher initial
concentration. In this way, the less favored pole is eventually
drained of its protein population. Starting from randomly distributed
cytoplasmic MinD and DivIVA, with average densities $1000~\mu m^{-1}$,
and $750~\mu m^{-1}$, respectively, an appropriate parameter set is:
$D_D=2.0~\mu m^2 s^{-1}$, $D_d=0$, $D_A=2.0~\mu m^2 s^{-1}$,
$D_a=2\times 10^{-4}~\mu m^2 s^{-1}$, $\lambda_D=10~s^{-1}$,
$\lambda_D'=0$, $\lambda_d=3.0~s^{-1}$, $\lambda_d'=0.001~\mu m$,
$\lambda_A=0.012~\mu m^2 s^{-1}$, $\lambda_A'=0$,
$\lambda_a=1.0~s^{-1}$, $\lambda_a'=1.2\times 10^{-4}~\mu m$. In this
case localization to a (randomly selected) pole occurs in about
$800~s$ (data not shown). Alternatively, replacing some of the above
parameters with $\lambda_d'=1\times 10^{-4}~\mu m$,
$\lambda_A=0.002~\mu m^2 s^{-1}$, $\lambda_a=1.6~s^{-1}$, and
$\lambda_a'=1.6\times 10^{-4}~\mu m$, unipolar localization occurs far
more rapidly, within about $40~s$.  A process analogous to this may be
happening in asymmetric division in sporulating {\em
B. subtilis} \cite{Ben-Yehuda1} (see below), although in that case
the protein SpoIIE may also play a crucial role \cite{Lucet}. 

\section{Discussion}

In this paper, we have proposed a mechanism for the symmetric or
asymmetric polar localization of proteins, relying on a combination of
geometric information and reaction--diffusion dynamics. This mechanism
could be particularly important for polar protein localization in
cells growing out from spores, where there are no pre--existing
division apparatus from prior division events. However, even in
vegetative cells, where the poles could be marked by the division
apparatus from prior divisions, the proposed mechanism could be useful
for increased reliability (and redundancy) in locating the poles. We
have focused particular attention on the MinCD/DivIVA system in {\em
B. subtilis}, where our results agree well with experiments on
outgrowing spores. We have also made a number of testable assumptions:
very low diffusion constants for membrane--bound MinCD/DivIVA; DivIVA
binds to the edges of MinD clusters; and membrane binding of MinD is
heavily influenced by geometric effects (possibly by the high
curvature of the cell poles restricting MinD membrane
polymerization). We should point out, however, that experiments in
round cell mutants have indicated that simple curvature of the cell
membrane is insufficient to elicit DivIVA targeting
\cite{Edwards2}. Nevertheless, this does not contradict our results:
according to our model, what is important is that there be a {\em
difference} in curvature: one part of the cell with low curvature, 
another part (the cell poles) with high curvature.

As we mentioned in the introduction there are several other regulatory
systems in bacteria with similar polar localization properties. For
example, the MCP chemotaxis receptor proteins (in {\em E. coli})
\cite{Maddock,Lybarger,Maki} and McpA (in {\em C. crescentus})
\cite{Alley} are both polar localized. Furthermore, filamentous
mutants in both cases show an intriguing pattern of localization: the
proteins are still polar localized, but in addition there is a
seemingly randomly distributed component along the filament length
between the two poles. This pattern is qualitatively reminiscent of
the pattern seen (and modeled in this paper) for MinCD/DivIVA in
filamentous {\em B. subtilis} mutants grown from spores
\cite{Harry,Hamoen}. Although, for the chemotaxis cases, it may simply
be that the proteins have saturated polar binding sites and are then
``spilling'' over into the remainder of the cell. However it is again
possible that there may be a geometric component to the localization,
similar to that proposed in this report. This may be even more likely
in the case of chemotactic proteins in outgrowing spores. Already the
chemotaxis protein McpB is known to be polar localized in vegetative
{\em B. subtilis} \cite{Kirby}.  It would be very interesting to study
its localization properties in outgrowing spores without pre--existing
division apparatus.  In addition, it has been shown that DivIVA
localizes to division sites both in {\em E. coli} and, unexpectedly,
in the fission yeast {\em Schizosaccharomyces pombe}
\cite{Edwards2}. However, this may again be due to binding to division
apparatus proteins (possibly actin in {\em S. pombe}).

Given that the geometric properties (high curvature) of the cell poles
may be important in influencing MinD binding, it is, of course,
important to understand how the cell shape is maintained. In {\em
B. subtilis}, cell elongation and maintenance of the cell cylinder is
achieved by Mbl--directed dispersed helical insertion of cell wall
material throughout the cylinder, while the hemispherical cell poles
are effectively inert \cite{Daniel}. In vegetative cells, where a
midcell septum is formed, it is plausible to see how the hemispherical
end is constructed and thus how overall cell shape
maintained. However, in spore outgrowth, this is not so obvious. {\em
B. subtilis} spores themselves are already ellipsoidally shaped, so
the initial state for spore germination already has the ends marked
via high curvature. However quite how this shape is produced in the
earlier sporulation event is rather unclear.

It is also intriguing to note that, although MinC and MinD are used by
both {\em B. subtilis} and {\em E. coli}, the two bacteria appear to
use entirely different physical mechanisms to exclude division from
polar sites. {\em E. coli} uses a spontaneous reaction--diffusion
instability, whereas geometric effects (reduced polar binding of MinD)
and associated reaction--diffusion kinetics, appear to be employed in
{\em B. subtilis}. Interestingly, it is actually a {\em reduced} rate
of MinD membrane binding at the cell poles which, in our model,
eventually leads to polar localization due to DivIVA
binding/stabilizing the edges of MinD membrane clusters. It would be
interesting to further study these mechanisms with their evolutionary
origin in mind, especially since both employ reaction-diffusion
dynamics. Although we believe that the mechanism outlined in this
paper provides a very plausible scenario for polar protein
localization, other mechanisms are still possible. For example, one
could imagine that the DivIVA makes the poles more attractive for
MinCD polymerization, perhaps by altering the structure of the MinCD
filaments so that they prefer a higher curvature region (see also
Ref.~\cite{Harry}). Given that the model described in this paper
provides several testable predictions, experiments should be readily
able to discriminate between these possible models.

One may also speculate on why {\em E. coli} and {\em B. subtilis} have
differing mechanisms. One possible explanation is that {\em
B. subtilis} must have extra flexibility in its cell division
regulation as, unlike {\em E. coli}, it sporulates when confronted
with harsh environmental conditions. In the process of forming a
spore, the division septum is relocated from midcell to one of the
cell poles. Hence {\em B. subtilis} must have the ability to undo its
polar division block. All that would be required is for another
protein to bind to DivIVA, displacing the MinCD, thus freeing up the
polar sites, as suggested in Ref.~\cite{Ben-Yehuda2}. A suitable
candidate for such a displacing protein has recently been shown to be
RacA \cite{Ben-Yehuda2}. The MinCD presumably then relocates to the
cell center, forcing the FtsZ ring to spiral outwards to the cell
poles, as observed in Ref.~\cite{Ben-Yehuda1}. As only a single pole
is then selected for spore formation, the cytokinetic proteins should
localize asymmetrically, possibly in a similar way to that
discussed above. 

Finally, we mention that the repertoire of division site positioning
is likely still not exhausted by the MinCDE and DivIVA systems
analyzed here and in previous modeling work
\cite{Howard1,Meinhardt,Kruse,Howard2}. {\em C. crescentus} is able to
divide asymmetrically with high precision even though it lacks all of
the MinCDE and DivIVA proteins \cite{Shapiro2}. How this is achieved
remains unknown and hints that there may be yet other mechanisms
capable of providing accurate positional information and hence of
regulating subcellular architecture.

\

\noindent
{\bf Acknowledgements}

\

\noindent 
This work was supported by The Royal Society.


\end{document}